\documentclass[conference]{IEEEtran}
\usepackage{amssymb}
\usepackage[space]{cite}
\usepackage{multirow}
\usepackage[acronym,nomain]{glossaries}
\usepackage{booktabs}
\usepackage{color,soul}
\usepackage{threeparttable}
\usepackage[per-mode=symbol]{siunitx}
\usepackage[inkscapelatex=false]{svg}
\usepackage{balance}
\usepackage{array}
\makeglossaries

\usepackage[printonlyused,withpage]{acronym}

\newcommand*{\sectionshift}{-0.2cm}

 \usepackage{tkz-euclide}
 \usetikzlibrary{arrows}
 \usepackage{setspace}
 \usepackage{bm}
 \usepackage{relsize}
 \usepackage[dvipsnames]{xcolor}
 \usepackage{psfrag}
 \usepackage{dblfloatfix}
 \usepackage{tabularx}
\usepackage{soul}
\usepackage{float}
 \usepackage{subfiles}
 \usepackage{epstopdf}
 \usepackage{verbatim}
 \usepackage{ifthen}
 \usepackage{pgfplots}
 \usepackage{pgfplotstable}
 \pgfplotsset{compat=1.18}
 \usepackage{tikzscale}
 \usepackage{amsfonts}
 \usepackage{tikz-3dplot}
 \usepackage{graphicx}
\usepackage{subfig}
\usepackage{footnote}
\makesavenoteenv{tabular}
\makesavenoteenv{table}

\usepackage{tikz}
\usepgfplotslibrary{fillbetween}
\usetikzlibrary{backgrounds}
\tikzstyle{load}   = [ultra thick,-latex]
\tikzstyle{stress} = [-latex]
\tikzstyle{dim}    = [latex-latex]
\tikzstyle{axis}   = [-latex,black!55]
\usetikzlibrary{arrows}
\usepackage{tikzscale}
\usepackage{tikz-3dplot}
\usetikzlibrary{3d}
\usetikzlibrary{shapes}
\usetikzlibrary {patterns,patterns.meta}
\usetikzlibrary{shapes.geometric}
\usetikzlibrary{angles,quotes}
\usetikzlibrary{fit}
\usepgfplotslibrary{groupplots}

\colorlet{veccol}{green!50!black}
\colorlet{projcol}{blue!70!black}
\colorlet{myblue}{blue!80!black}
\colorlet{myred}{red!90!black}
\colorlet{mydarkblue}{blue!50!black}
\definecolor{green(pigment)}{rgb}{0.0, 0.65, 0.31}
\definecolor{LISmagenta}{HTML}{e312ff}
\definecolor{LISpurple}{HTML}{000576}

\tikzset{>=latex} 
\tikzstyle{proj}=[projcol!80,line width=0.08] 
\tikzstyle{area}=[draw=veccol,fill=veccol!80,fill opacity=0.6]
\tikzstyle{vector}=[-stealth,myblue,thick,line cap=round]
\tikzstyle{unit vector}=[->,veccol,thick,line cap=round]
\tikzstyle{dark unit vector}=[unit vector,veccol!70!black]
\usetikzlibrary{angles,quotes} 

\newsavebox{\myimage}

\definecolor{cadmiumgreen}{rgb}{0.0, 0.42, 0.24}
\definecolor{cobalt}{rgb}{0.0, 0.28, 0.67}

\DeclareMathAlphabet{\mathsfbr}{OT1}{cmss}{m}{n}
\SetMathAlphabet{\mathsfbr}{bold}{OT1}{cmss}{bx}{n}
\DeclareRobustCommand{\msf}[1]{%
	\ifcat\noexpand#1\relax\msfgreek{#1}\else\mathsfbr{#1}\fi
}

\renewcommand{\vec}[1]{\ensuremath{\boldsymbol{#1}}}
\newcommand{\mrm}[1]{\ensuremath{\mathrm{#1}}}

\newacronym{2d}{2D}{two-dimensional}
\newacronym{3d}{3D}{three-dimensional}
\newacronym{5g}{5G}{5th-generation}
\newacronym{pnt}{PNT}{positioning, navigation and timing}
\newacronym{pa}{PA}{physical anchor}
\newacronym{va}{VA}{virtual anchor}
\newacronym{bs}{BS}{base station}
\newacronym{mpc}{MPC}{multipath component}
\newacronym{slam}{SLAM}{simultaneous localization and mapping}
\newacronym{fa}{FA}{false alarm}

\newacronym{afe}{AFE}{analog front-end}
\newacronym{pcb}{PCB}{printed circuit board}
\newacronym{tx}{TX}{transmitter}
\newacronym{rx}{RX}{receiver}
\newacronym{lna}{LNA}{low noise amplifier}
\newacronym{baw}{BAW}{Bulk-Acoustic Wave}

\newacronym{adc}{ADC}{analog-to-digital converter}
\newacronym{ai}{AI}{Artificial Intelligence}
\newacronym{aoa}{AoA}{angle-of-arrival}
\newacronym{aod}{AoD}{angle-of-departure}
\newacronym{ap}{AP}{access point}
\newacronym{asic}{ASIC}{application specific integrated circuit}
\newacronym{asip}{ASIP}{application specific integrated processor}
\newacronym{awgn}{AWGN}{additive white Gaussian noise}

\newacronym{bp}{BP}{belief propagation}

\newacronym{ce}{CE}{channel estimation}
\newacronym{cw}{CW}{continuous wave}
\newacronym{cdf}{CDF}{cumulative distribution function}
\newacronym{ced}{CED}{cumulative energy density}
\newacronym{clk}{CLK}{clock}
\newacronym{csi}{CSI}{channel state information}
\newacronym{csp}{CSP}{contact service point}
\newacronym{cpu}{CPU}{central processing unit}
\newacronym{crlb}{CRLB}{Cramer-Rao lower bound}
\newacronym{dc}{DC}{direct current}
\newacronym{dfe}{DFE}{digital front-end}
\newacronym{dfg}{DFG}{data-flow graph}
\newacronym{dlp}{DLP}{downlink pilot}
\newacronym{dbb}{DBB}{digital baseband}
\newacronym{dld}{DLD}{downlink data}
\newacronym{dof}{DoF}{degrees of freedom}
\newacronym{dmimo}{D-MIMO}{distributed massive MIMO}
\newacronym{ram}{RAM}{random-access memory}
\newacronym{ecc}{ECC}{elliptic curve cryptography}
\newacronym{ecdlp}{ECDLP}{elliptic curve discrete logarithm problem}
\newacronym{ecsp}{ECSP}{edge computing service point}
\newacronym{eirp}{EIRP}{equivalent isotropically radiated power}
\newacronym{elaa}{ELAA}{extremely large aperture array}
\newacronym{em}{EM}{electromagnetic}
\newacronym{en}{EN}{energy neutral}
\newacronym{e2e}{E2E}{End-to-End}
\newacronym{erfc}{erfc}{complementary error function}
\newacronym{gpio}{GPIO}{general-purpose input-output}
\newacronym{pe}{PE}{processing engine}
\newacronym{rtl}{RTL}{register transfer level}
\newacronym{hls}{HLS}{high-level synthesis}
\newacronym{fft}{FFT}{fast fourier transform}
\newacronym{fhss}{FHSS}{frequency hopping spread spectrum}
\newacronym{fl}{FL}{federated learning}
\newacronym{fpga}{FPGA}{field programmable gate array}
\newacronym{gdpr}{GDPR}{general data protection regulation}
\newacronym{ic}{IC}{integrated circuit}
\newacronym{ioe}{IoE}{Internet of Everything}
\newacronym{iot}{IoT}{Internet of Things}
\newacronym{id}{ID}{information decoding}
\newacronym{kpi}{KPI}{key performance indicator}
\newacronym{lca}{LCA}{life cycle assessment}
\newacronym{ls}{LS}{least squares}
\newacronym{lis}{LIS}{large intelligent surface}
\newacronym{liion}{Li-ion}{lithium-ion}
\newacronym{los}{LoS}{line-of-sight}
\newacronym{lut}{LUT}{look-up table}
\newacronym{nlos}{NLoS}{non-line-of-sight}
\newacronym{olos}{OLoS}{obstructed-line-of-sight}
\newacronym{lmo}{LMO}{lithium ion manganese oxide}
\newacronym{mami}{MaMi}{massive MIMO}
\newacronym{mf}{MF}{matched filter}
\newacronym{ml}{ML}{machine learning}
\newacronym{mmse}{MMSE}{minimum mean square error}
\newacronym{mmwave}{mmWave}{millimeter wave}
\newacronym{mimo}{MIMO}{multiple-input multiple-output}
\newacronym{miso}{MISO}{multiple-input single-output}

\newacronym{mrc}{MRC}{maximum ratio combining}
\newacronym{mrt}{MRT}{maximum ratio transmission}
\newacronym{ofdm}{OFDM}{orthogonal frequency-division multiplexing}
\newacronym{ota}{OTA}{over-the-air}
\newacronym{ii}{II}{Initiation interval}
\newacronym{nb}{NB}{narrowband}
\newacronym{nr}{NR}{New Radio}
\newacronym{pda}{PDA}{probabilistic data association}
\newacronym{pdp}{PDP}{power delay profile}
\newacronym{pdf}{PDF}{probability density function}
\newacronym{spa}{SPA}{sum-product algorithm}

\newacronym{per}{PER}{packet error rate}
\newacronym{pet}{PET}{privacy enhancing technology}
\newacronym{pki}{PKI}{public key infrastucture}
\newacronym{pc}{PC}{pilot count}
\newacronym{pnn}{PNN}{probabilistic neural network}
\newacronym{plhf}{PLHF}{pseudo-likelihood function}
\newacronym{pssh}{PSSH}{parallel secure shell}
\newacronym{qrrls}{QR-RLS}{QR decomposition based recursive least squares}
\newacronym{re}{RE}{radio element}
\newacronym{rf}{RF}{radio frequency}
\newacronym{rfid}{RFID}{radio frequency identification}
\newacronym{ris}{RIS}{reflective intelligent surface}
\newacronym{rmse}{RMSE}{root mean square error}
\newacronym{rls}{RLS}{recursive least squares}
\newacronym{rss}{RSS}{received signal strength}
\newacronym{rssi}{RSSI}{received signal strength indicator}
\newacronym{rw}{RW}{RadioWeaves}
\newacronym{sa}{SA}{synchronization anchor}
\newacronym{sdg}{SDG}{Sustainable Development Goal}
\newacronym{sdn}{SDN}{software-defined network}
\newacronym{sdr}{SDR}{software-defined radio}

\newacronym{se}{SE}{spectral efficiency}
\newacronym{sinr}{SINR}{signal-to-interference-plus-noise ratio}
\newacronym{sir}{SIR}{signal-to-interference ratio}
\newacronym{siso}{SISO}{single-input single-output}
\newacronym{snr}{SNR}{signal-to-noise ratio}
\newacronym{srls}{SRLS}{standard recursive least squares}
\newacronym{svd}{SVD}{singular value decomposition}
\newacronym{sp}{SP}{service point}
\newacronym{sram}{SRAM}{static random-access memory}
\newacronym{tdd}{TDD}{time division duplexing}
\newacronym{tdoa}{TDOA}{time-difference-of-arrival}
\newacronym{toa}{TOA}{time-of-arrival}
\newacronym{ue}{UE}{user equipment}
\newacronym{uhf}{UHF}{ultra high frequency}
\newacronym{ula}{ULA}{uniform linear array}
\newacronym{ulp}{ULP}{uplink pilot}
\newacronym{ul}{UL}{uplink}
\newacronym{uld}{ULD}{uplink data}
\newacronym{upa}{UPA}{uniform planar array}
\newacronym{ura}{URA}{uniform rectangular array}
\newacronym{uwb}{UWB}{ultrawideband}
\newacronym{wb}{WB}{wideband}
\newacronym{wpt}{WPT}{wireless power transfer}
\newacronym{zf}{ZF}{zero-forcing}

\newacronym{lpt}{LPT}{laser power transfer}
\newacronym{rfpt}{RFPT}{radio frequency power transfer}
\newacronym{ipt}{IPT}{inductive power transfer}
\newacronym{cpt}{CPT}{capacitive power transfer}
\newacronym{apt}{APT}{acoustic power transfer}
\newacronym{cfo}{CFO}{carrier frequency offset}
\newacronym{lo}{LO}{local oscillator}
\newacronym{if}{IF}{intermediate frequency}
\newacronym{duc}{DUC}{digital up conversion}
\newacronym{ddc}{DDC}{digital down conversion}
\newacronym{dsp}{DSP}{digital signal processing}
\newacronym{vco}{VCO}{voltage-controlled oscillator}
\newacronym{pll}{PLL}{phase-locked loop}
\newacronym{rt}{RT}{ray-tracing}
\newacronym{dac}{DAC}{digital-to-analog converter}
\newacronym{de}{DE}{drain efficiency}
\newacronym{pae}{PAE}{power-added efficiency}
\newacronym{sar}{SAR}{specific absorption rate}
\newacronym{mppt}{MPPT}{maximum power point tracking}
\newacronym{isi}{ISI}{intersymbol interference}
\newacronym{pps}{PPS}{pulse per second}
\newacronym{ps}{PS}{processing system}
\newacronym{pl}{PL}{programmable logic}
\newacronym{icnirp}{ICNIRP}{International Commission on Non-Ionizing Radiation Protection}
\newacronym{fd}{FD}{Front-haul distance}
\newacronym{wd}{WD}{Wireless distance}
\newacronym{ntp}{NTP}{Network Time Protocol}
\newacronym{ptp}{PTP}{Precision Time Protocol}
\newacronym{wr}{WR}{White Rabbit}
\newacronym{mm2s}{MM2S}{memory mapped to stream}
\newacronym{s2mm}{S2MM}{stream to memory mapped}
\newacronym{ber}{BER}{bit error rate}
\newacronym{mu}{MU}{multiplication unit}
\newacronym{gp}{GP}{general purpose}
\newacronym{hp}{HP}{high performance}
\newacronym{ids}{IDS}{indoor dense space}
\newacronym{dma}{DMA}{direct memory access}
\newacronym{qam}{QAM}{Quadrature amplitude modulation}
\newacronym{qpsk}{QPSK}{Quadrature  phase shift keying}
\newacronym{sd}{SD}{sphere decoding}
\newacronym{xr}{XR}{extended reality}
\newacronym{xlmimo}{XL-MIMO}{extremely large scale MIMO}

\usetikzlibrary{external}
\IEEEoverridecommandlockouts

\usepackage[skip=6pt]{caption}

\begin{document}
\bstctlcite{IEEEexample:BSTcontrol}

\title{
A Scalable 256-Antenna Distributed MIMO Testbed with Real-Time Fully Digital Beamforming}
\vspace{-1mm}

\author{
\IEEEauthorblockN {Dumitra Iancu\IEEEauthorrefmark{1},
Vilgot Snygg\IEEEauthorrefmark{1},
Sijia Cheng\IEEEauthorrefmark{1},
Lina Tinnerberg\IEEEauthorrefmark{1},
Mikael Henriksson\IEEEauthorrefmark{2},
Emil Bergman\IEEEauthorrefmark{1}, \\
Anders J Johansson\IEEEauthorrefmark{1}, 
Baktash Behmanesh\IEEEauthorrefmark{1}, 
Ove Edfors\IEEEauthorrefmark{1}, 
Liang Liu\IEEEauthorrefmark{1}
}
\\ \vspace{-2mm}
\IEEEauthorblockA{\IEEEauthorrefmark{1} Department of Electrical and Information Technology, Lund University, Sweden}
\IEEEauthorblockA{\IEEEauthorrefmark{2} Department of Electrical Engineering, Linköping University, Sweden}

Email: \{firstname.lastname\}@eit.lth.se, @liu.se
\thanks{This work is funded by the Swedish Foundation for Strategic Research (SSF) project CHI19-0001: Large Intelligent Surfaces – Architecture and Hardware and the strategic research area ELLIIT, through the LISA research infrastructure initiative.}
}
 
\maketitle

\begin{abstract}
    \Gls{dmimo} is a promising technology for future generation wireless systems as it takes advantage of  both an increased array aperture and a decentralized processing architecture and topology. In order to truly understand the possibilities and limitations of these approaches in real scenarios, practical realization of testbeds is an essential step in the technology advancement. This work presents the Lund University Large Intelligent Surface testbed --- LuLIS, that can operate up to 256 coherent \gls{rf} chains using 16 AMD Zynq UltraScale+ RFSoC ZCU216 evaluation boards acting as distributed processing nodes. Real-time processing is facilitated by acceleration and distribution of MIMO processing algorithms on the FPGA fabric of the boards. The system is easily scalable, as increasing the number of antennas is done in multiples of 16 by adding more RFSoCs, which also implies  addition of another processing node. The design  allows up-scaling without hardware redesign, introduction of large latencies or data transfer overhead. The testbed is flexible in terms of deployment, with options of fully distributing the nodes (as in \Gls{dmimo}) or co-locating them (as in more traditional Massive MIMO). A detailed description of the implementation of the testbed is presented and initial results are shown for an \gls{ul} transmission from four single-antenna \glspl{ue} to 64, 128 and 256 base-station antennas for both a co-located and a distributed scenario.

\end{abstract}

\glsresetall
\IEEEpeerreviewmaketitle
\vspace{\sectionshift}

 \vspace{3mm}
\section {Introduction}
\label{sec:Introduction}




\Gls{dmimo} is a promising approach in developing next-generation wireless systems as it combines the advantages of large-scale antenna arrays and decentralized node architectures. Increasing the number of antennas at the \gls{bs} increases the number of spatial degrees of freedom, thus achieving high spectral efficiency and robust connectivity across spatially-multiplexed users, using the same time-frequency resource. Distributed architectures can mitigate the bottlenecks that typically come with centralized processing, 
while also providing better coverage, spatial diversity and user separation. Implementing systems that exploit the large array aperture expected in \gls{xlmimo} \cite{xlmimo}, or \glspl{lis} \cite{ogLIS}, while maintaining a distributed processing architecture, introduces several challenges, such as coherently processing large volumes of data, proportional to the number of antennas. However, 
real-life realization of testbeds is a pivotal step in validating and understanding hardware limitations of envisioned, theoretical systems \cite{testbeds}, and is a natural step toward continued improvement of existing technologies.

Notable efforts in implementing MIMO testbeds with a massive amount of antennas \cite{Lumami, huawei} need to be mentioned. However, these systems rely on centralized processing schemes, which come with several drawbacks \cite{studer2018}: (i) centrally performing massive amounts of computations scaling with the number of antennas; (ii) high inter-connection throughput from the antennas to a \gls{cpu}; (iii) limited scalability, as adding antennas often requires hardware redesign. Decentralized architectures that can alleviate some of these problems have been explored \cite{8960442, 8335613, 10256071}, however, to the best of our knowledge, the presented testbed is the first large-scale real-time end-to-end \Gls{dmimo} implementation.

This work introduces the  Lund University Large Intelligent Surface testbed (LuLIS), with 256 antennas and 256 fully-parallel \gls{rf} chains, with distributed base-band processing, scalable, flexible and performs fully coherent real-time digital beamforming.



\begin{figure}[t]
    \centering 
        \includegraphics[scale=2.3]{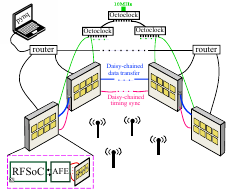}
    \caption{Depiction of the distributed, scalable, antenna panel based design}
    \label{fig:high-level}
\end{figure}
 

\section{System model and testbed design}
\label{sec:TestbedDesign}
We consider a distributed panel-based MIMO architecture, consisting of $J=16$ panels deployed in an area of interest and acting as the \gls{bs}. Each antenna panel has $M=16$ antenna ports, arranged as an 4x2 rectangular array of eight dual-polarized patch antennas, and is equipped with local processing units that are interconnected and exchanging data through an Ethernet-based fronthaul network. The system is \gls{ofdm}-based and is serving $K$ single-antenna users in the \gls{ul}. 

\subsection{Distributed panel-based design}
\label{subsec:Distributed panel-based design}

   

Fig. \ref{fig:high-level} shows a conceptual depiction of the testbed. The distributed RFSoC-based architecture, with near-antenna processing and daisy-chain topology, has the following features:
\begin{enumerate}
    \item Distributed, custom processing. The \gls{dbb} processing  is co-designed with the system topology and parts of the MIMO equalization is performed on local processing units, near the antennas.  We leverage \gls{fpga}-based processing to accelerate the \gls{dbb} operations and heavily parallelize the processing, facilitating real-time communications for applications such as \gls{xr}.  All units are synchronized to perform coherent processing.
    \item Constant data interconnection rates between computing nodes. The need for sending the full channel matrix to a centralized processor disappears, as data is partially processed locally. Due to the distribution of the compute, the transfer data rates are constant throughout the entire processing chain, scaling solely with the number of users, $K$. This reduces the latency from the fronthaul as less data is moved between the processing nodes. 
    \item Scalability and flexibility. The testbed is a \gls{sdr}-based system designed to be modular, as essentially identical operations facilitate the addition of another local node in the chain without doing any redesign. Moreover, the testbed is software-configurable and flexible in terms of deployment, with options such as co-locating all antennas or fully distributing them.  

\end{enumerate}

 Each panel incorporates the antenna array, and an \gls{afe} interfacing with a AMD Zynq UltraScale+ RFSoC ZCU216 evaluation board. The development platform features an integrated chip with 16 high speed \gls{rf} \gls{adc}/\gls{dac}, \gls{fpga} \gls{pl} and a \gls{ps} consisting of multiple ARM cores, making it suitable for flexible \gls{sdr}-based applications.
The fronthaul infrastructure, which transmits the data between the panels, is implemented through 25Gb Ethernet ports which are connected in a daisy-chain topology.
A computer acting as a central controller is connected to all the panels through a chain of routers. It manages interfacing and configuration of the testbed, deploys FPGA bitstreams, and provides real-time visualization for the equalized received user data. This is done leveraging the PYNQ framework \cite{pynq} run through \gls{pssh}.

\begin{figure*}[!t]
    \centering 
        \includegraphics[scale=0.54]{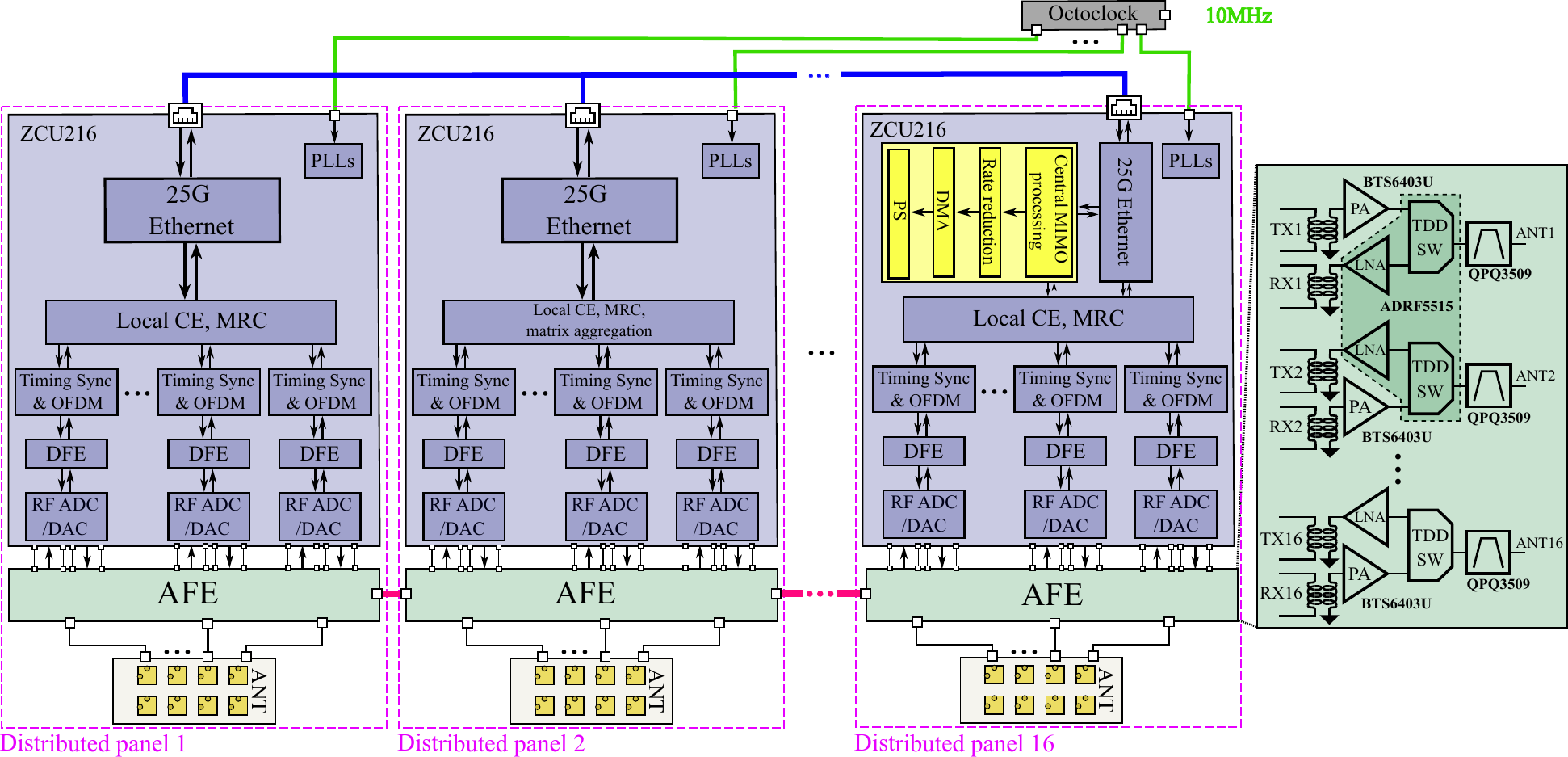}
    \caption{ Block diagram of the processing capabilities of the panels showing only the data path.}
    \label{fig:bdd}
\end{figure*}


Fig. \ref{fig:bdd} shows a detailed block diagram of the system partitioning across processing nodes. Each panel features 16 parallel receiver chains. For each chain, each antenna pin is connected to the \gls{afe}, where the incoming signals are filtered and amplified. Then they are digitized through the on-chip data converters and later processed by the digital blocks implemented on the \gls{pl} fabric of the ZYNQ chip. In the following sections we will present in detail all the subsystems and highlight the implementation features of the testbed.

\subsection{Antenna array}
\label{subsec:Antenna array}

Each antenna panel hosts a 2x4 planar array of antenna elements with $\lambda/2$ spacing between the elements in both directions. The elements are dimensioned to have a center frequency at 3.84 GHz with 100 MHz bandwidth. A stacked architecture with square patches printed on two substrates separated by a 4.5~mm air gap was chosen to achieve high bandwidth while maintaining a compact form factor. The lower patch is constructed like a classical microstrip antenna with dimension 18.27 x 18.27 mm placed above a ground plane. It is driven by two feed pins, each placed 6.40~mm from the center towards the edge, at a 90 degree angle from each other to allow for horizontal and vertical polarization. The driven element is shorted to its ground plane using a 0.2~mm via in the center to improve isolation between the two polarization modes. The sides of the passive upper patch were made 3\% larger than the sides of the driven element to increase bandwidth and improve matching to the AFE, resulting in the dimension and 18.81 x 18.81~mm. Both substrates are made out of Rogers ro4350b due to its availability, favorable dielectric constant, and low losses. 

\subsection{Analog design}
\label{subsec:Analog design}

Each antenna array is fed via a custom-built \gls{afe} that supports up to 16 parallel \gls{tx} and \gls{rx} channels. A simplified block diagram of the \gls{afe} is shown in Fig. \ref{fig:bdd}. The \gls{afe} supports \gls{tdd} operation by utilizing the dual-channel receiver front-end chip, ADRF5515 from Analog Devices. This chip includes a Single-Pole-Double-Throw (SPDT) \gls{tdd} switch per channel. The on-chip, channel-to-channel isolation is 45 dB. The \gls{tdd} mode is controlled by the RFSoC via the \gls{gpio}. The ADRF5515 chip also includes one \gls{lna} per channel, offering 32 dB of maximum gain and a noise figure of around 1.1 dB at 3.84 GHz. A single-ended to differential balun is included at the output of each \gls{lna} on the PCB to minimize the effect of the common-mode noise and separate the bias domains to drive the differential input ADCs on the RFSoC.

On the \gls{tx} side, the BTS6403U driver amplifier from NXP Semiconductors is utilized as the power amplifier, providing a power gain of around 38 dB at 3.84 GHz. The differential signals from the DACs on the RFSoC are converted to single-ended signal close to the input of the BTS6403U. A BAW bandpass filter, QPQ3509 from Qorvo, is utilized between the \gls{tdd} switches and antennas to mitigate out-of-band transmission in \gls{tx} mode and limit the in-band noise in the \gls{rx} mode. The filter has a passband bandwidth of 280 MHz around 3.84 GHz with a steep roll-off behavior and a typical in-band insertion loss of around 2.2 dB.

\begin{figure}[t]
    \centering 
        \includegraphics[scale=0.045]{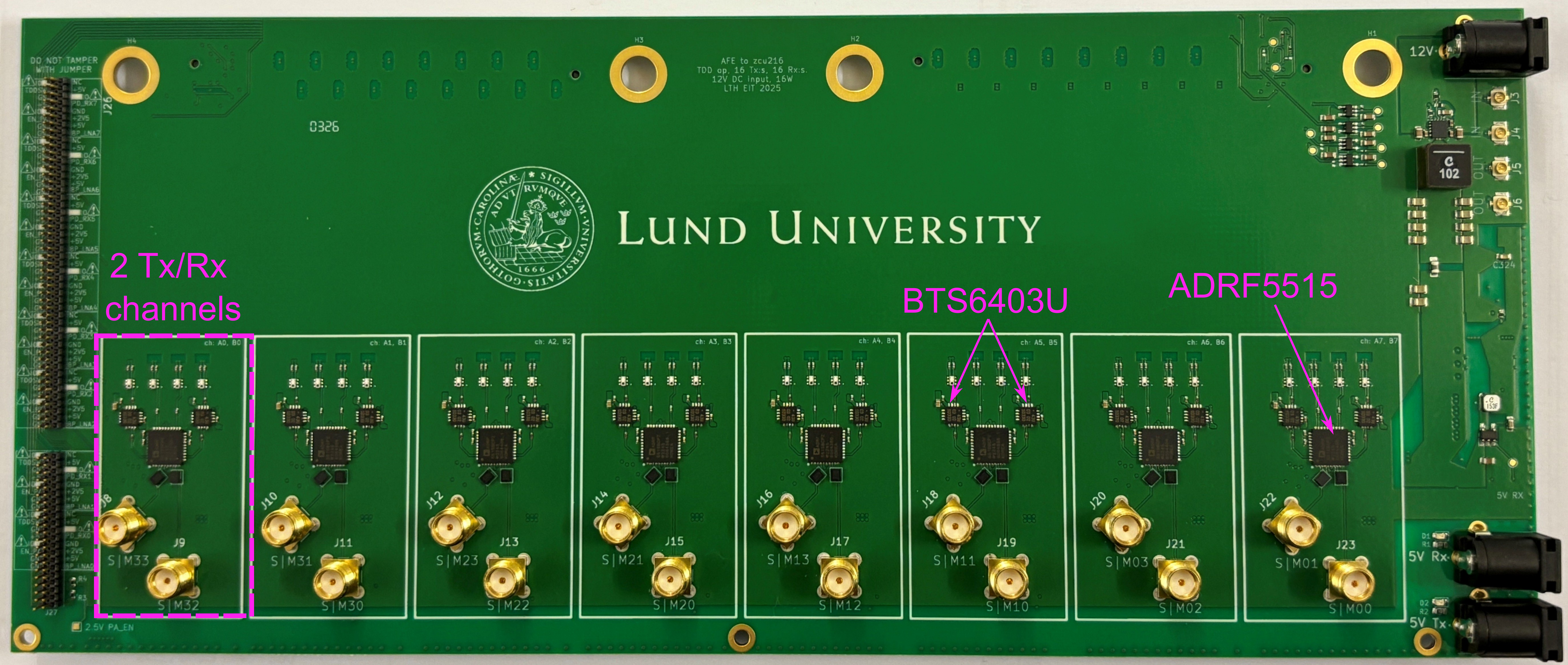}

    \caption{Fabricated PCB of the 16 channel \gls{afe}.
    }
    \label{fig:afe_pic}
\end{figure}

The \gls{afe} is powered using an external 12V power supply, which is regulated down to two separate 5V supplies on the board for \gls{rx} and \gls{tx} circuits, respectively. This separation protects high-sensitivity \gls{rx} channels from the noise of strong \gls{tx} signals that could be coupled through the supplies. Furthermore, all individual supply rails have additional filtering close to the input supply pins on their respective chip. Fig. \ref{fig:afe_pic} shows a picture of the fabricated 16-channel \gls{afe} PCB.

\subsection{Digital design}
\label{subsec:Digital design}

Each ZCU216 board implements 16 parallel \gls{dbb} chains on the \gls{pl}, each serving one \gls{bs} antenna input, as shown in Fig. \ref{fig:bdd}. The digital front-end is implemented through the Zynq Ultrascale+ RF Data Converter IP \cite{dataconv}, with an integrated mixer and decimator acting as the \gls{dfe} and is used and configured to enable and synchronize all the \glspl{adc}. \Gls{ofdm} functionality, such as \gls{fft}, removal of cyclic prefixes and guardband subcarriers as well as timing synchronization is implemented as the next processing step in the chain. The data streams from all 16 chains are aggregated, local channel estimation and the local Gram matrix are computed. \Gls{mrc} equalization is implemented, following the algorithmic design in \cite{tinnerberg2025spectrumefficiencyprocessinglatency}.  We reiterate here the equation of how the computations are distributed for the benefit of the reader. The local Gram matrix and \gls{mrc} post-processing vector are computed on each local processing node, and accumulated in the processing chain, according to

\begin{equation}\label{eq:mrc_post}
  \begin{aligned}
\boldsymbol{z}_\mathrm{MRC} &= \sum_{j=1}^{J}\boldsymbol{H}_j^\mathrm{H}\boldsymbol{y}_j
= \sum_{j=1}^{J}\vec{z}_{\mathrm{MRC},j}
\end{aligned}
\end{equation} 

\begin{equation}\label{eq:zf_post}
\begin{aligned}
\boldsymbol{G}=\sum_{j=1}^{J}\boldsymbol{H}_j^\mrm{H}\boldsymbol{H}_j= \sum_{j=1}^{J}\boldsymbol{G}_j
\end{aligned}
\end{equation}
where  $\vec{z}_{\mathrm{MRC},j}$ is the local MRC-post-processed vector, $\boldsymbol{y}_j$ is the received vector, $\boldsymbol{H}_j$ is the local channel matrix and $\boldsymbol{G}_j$ is the local Gram matrix at panel $j$. The Gram matrix will be used later for \gls{zf} decoding.

The computed partial sum term for all subcarriers is packaged and sent through the fronthaul network. The local computations decrease the necessary bandwidth compared to if the raw channel samples were transmitted, as the data now scales solely with the amount of users, $K$, instead of also scaling with the amount of antennas, $M$. This design choice also keeps the amount of data sent between each panel pair constant, instead of increasing after every connection, thus making the design for the fronthaul network much simpler. Each middle node is now identical, making the design easily scalable as described in Section \ref{subsec:Distributed panel-based design}. The bandwidth utilized, between every panel, is  1.73Gb/s for 4 users and with the frame structure detailed in Section \ref{subsec:General parameters}.



All digital blocks are performing signal handshakes following AXIS/AXI protocols. Table \ref{table:fpga-util} shows the utilized silicon resources on the \gls{fpga}, being around a third of the total resources. This leaves ample room for adding downlink functionality, and possibly other wireless-based services.

\vspace{4mm}
\begin{table}[h]
\centering
\caption{Utilization table for the digital blocks implemented on the \gls{fpga} for uplink transmission}
\resizebox{0.92\columnwidth}{!}{%
\begin{tabular}{@{}llllll@{}}
\toprule
\multicolumn{1}{l|}{Node} & Resources{[}\%{]} & LUT & DSP & RAM & FF \\ \midrule
\multicolumn{2}{l}{Local}                     & 26  & 31  & 35  & 17 \\
\multicolumn{2}{l}{Central}                   & 27  & 31  & 42  & 17 \\ \bottomrule
\end{tabular}%
}
\label{table:fpga-util}
\end{table}

\label{sec:digitaldesign}

\subsection{Synchronization }
\label{subsec:Synchronization}

\subsubsection{Frequency synchronization}
\label{subsec:Frequency synchronization}

The distributed nodes are frequency synchronized through a 10MHz reference clock that is distributed with three CDA-2990 octoclocks connected in a tree structure with high-frequency SMA cables. The reference clock is fed into the CLK104 add-on card \cite{clk104} with \glspl{pll} that later provide the source clocks for the on-board data converters.

\subsubsection{Timing synchronization}
\label{subsec:Timing synchronization}

The \gls{ue} is sending a timing synchronization signal through the \glspl{gpio} to all the \gls{bs} nodes. The timing synchronization signal is propagated through UFL to SMA cables between the \gls{bs} nodes through the same daisy-chain topology as for data transfer. The timing delay between the transmitters and receivers has been measured on the \gls{fpga} and then the synchronization signal delayed accordingly, through a python-configurable digital block. This allows for fine tuning through software, should different cables needed be used for different scenarios.

\subsection{User equipment}
\label{subsec:User equipment}

A 17th ZCU216 RFSoC with an attached \gls{afe} configured in transmit mode and connected to 4 omnidirectional antennas is used to simulate the \glspl{ue}. On the \gls{pl}, 4 memories are sending known pilots alongside synthetic \gls{ofdm} data symbols.
 

\subsection{Software central control}
\label{subsec:Software central control}


The digital blocks implemented on the \gls{fpga} are software-configurable through the \gls{ps}, which runs the PYNQ framework. The computer accesses and programs each node, with its own assigned IP address, via \gls{pssh}, with convenient checkpoints designed for debugging purposes. Using this set up it is easy to both add or remove nodes. 

The PYNQ framework is also used to provide real-time visualization of the received data, which is useful both during the debugging phase and when deploying the test bed in different environments. The data visualization is implemented by sending the post-processed vectors and the Gram matrix to the \gls{ps} through the \gls{dma}. To match the throughput of the \gls{dma}, a rate reduction block, which drops frames, is implemented in the \gls{pl}. 

In software, the possibility of calculating the \gls{zf} vector, using the post-processed vectors and the Gram matrix, is also available, allowing the performance achiveved with the testbed to be showcased.



\section{System specifications}
\label{sec:SystemSpecifications}
\subsection{General parameters}
\label{subsec:General parameters}

 The system specifications in Table \ref{table:specs} have been chosen to keep the cost of the testbed development relatively low and to fit the 5G numerology.

\vspace{4mm} 
\begin{table}[ht]
\centering
\caption{System parameters}
\resizebox{0.98\columnwidth}{!}{%
\begin{tabular}{@{}ll@{}}
\toprule
Parameter                                          & Value \\ \midrule
Center frequency {[}GHz{]}                         & 3.84  \\
Channel bandwidth {[}MHz{]}                        & 50    \\
ADC/DAC sampling rate {[}GS/s{]}                   & 2.457/4.915  \\
Baseband sampling rate {[}MS/s{]}                   & 61.44 \\
FFT size {[}number of samples{]}                    & 1024  \\
OFDM symbol duration {[}us{]}                      & 16.67 \\
Subcarrier spacing {[}kHz{]}                       & 60    \\
CP length {[}number of samples{]}                  & 144   \\ 
\gls{fpga} clock frequency{[}MHz{]}                 & 153.6 \\
Number of antennas/\gls{rf} chains per panel       & 16 \\
Number of panels (J)         & 16 \\
\bottomrule
\end{tabular}%
}
\label{table:specs}
\end{table}

A frame is composed of 7 OFDM symbols, \gls{ulp}, \gls{uld}, \gls{uld}, GUARD, \gls{dlp}, \gls{dld}, GUARD.
The guard symbol is chosen such that it allows for \gls{tdd}.

\subsection{Latency analysis}
\label{subsec:Latency analysis}

To evaluate the real-time performance of the realized system, the latency of the system blocks, depicted in Fig. \ref{fig:bdd}, was measured. In this analysis, the latency considered is the time from when the first input sample is provided to the analyzed block to when the processed sample is ready at the output. 

The main contribution of the processing latency are the digital blocks, and we will therefore consider the \gls{afe} and  RF Data Converter IP latency negligible. In Table \ref{table:latency} the latency of the different digital blocks and the full system latency, with 16 panels, is presented. There is no latency measurement for the central processing block as for the \gls{mrc} implementation there is no extra central processing needed beyond the data aggregation.



\begin{figure*}[!t]
\includegraphics[width=1\textwidth]{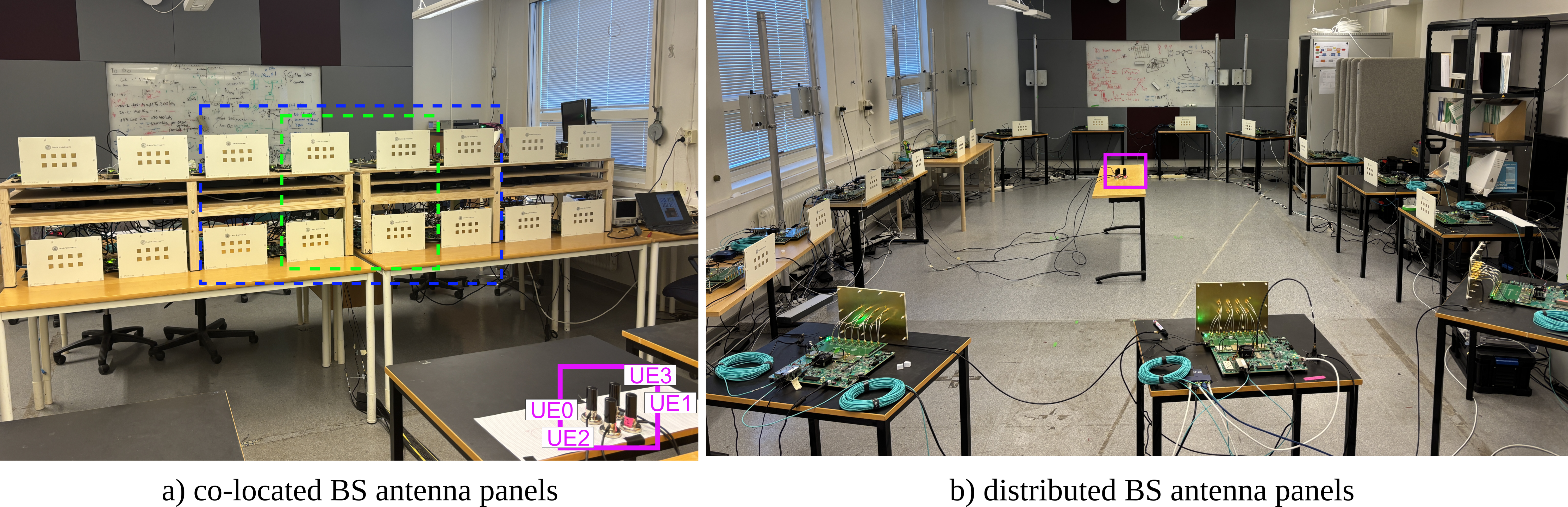}
\caption{The 16-panel testbed deployed in two different scenarios. The users, signaled with the magenta square, are tightly placed in a square at around $1.7$~m for the co-located scenario and at around $2$~m for the distributed scenario from the antenna array.}
\label{fig:testbed}
\end{figure*}

\vspace{4mm} 
\begin{table}[ht]
\centering
\caption{Measured latency }
\resizebox{0.98\columnwidth}{!}{%
\begin{tabular}{@{}lll@{}}
\toprule
Digital Block                                    &   Latency [cc]   & Latency [$\mu s$] \\ \midrule
Timing sync \& OFDM  &  6357   & 41.39  \\
Local CE \& MRC  &  21   &  0.14 \\
Ethernet \& Data Aggregate &   388  &  2.5 \\
System latency with $J=16$ &   12 198  &  79.41 \\

\bottomrule
\end{tabular}%
}
\label{table:latency}
\end{table}

The total processing latency in Table \ref{table:latency} is calculated using: 

\begin{align}
    \tau = \tau_{\mathrm{local}} + \tau_{\mathrm{transfer}}(J-1).
\end{align}
where $\tau_{\mathrm{local}}$ is the latency for both the Timing sync \& OFDM block, and the Local CE \& MRC block, $\tau_{\mathrm{transfer}}$ is the latency of the Ethernet fronthaul and the data aggregate, and $J$ is the number of panels. 

One important thing to note is that further latency optimizations of the design, especially the fronthaul, would be possible if needed for a latency critical application.

\section{Initial results}
\label{sec:InitialResults}

\subsection{Antenna array and \gls{afe}}
\label{subsec:Antenna and afe}

Characterization of one of the antenna panels showed that the reflection losses across all its individual elements were less than -10 dB at the center frequency of 3.84 GHz, with the worst case reflection loss of -10.3 dB. Compared to simulations, the matching notch is shifted up in frequency. The fabricated \gls{afe} achieved a worst case channel to channel crosstalk of -44.5 dBc across all 16 channels, indicating that the design is limited by the on-chip isolation of the ADRF5515 (-45 dB). In addition, the maximum output power for all 16 \glspl{tx} was measured during a single tone test with the full \gls{dac} voltage swing. An average output power of 18.7 dBm was achieved, with the minimum being 16.2 dBm.

\subsection{Uplink transmission test}
\label{subsec:Uplink transmission test}

Fig. \ref{fig:testbed} shows the testbed both in the co-located case, and distributed within the area of interest. 
\begin{figure}[ht]
	\centering
	\captionsetup[subfigure]{oneside,margin={0.85cm,0cm}}
	\hspace{-2mm}\subfloat[]{\hspace{2mm}\scalebox{0.45}{\includegraphics[scale=1]{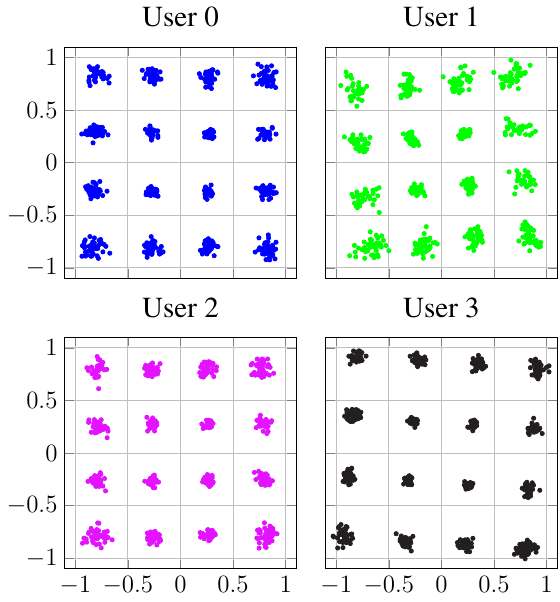}}\label{subfig:constellation-colocated}}    
	\captionsetup[subfigure]{oneside,margin={0.85cm,0cm}}
    \hspace{-8.5mm}\subfloat[]{\hspace{8mm}\scalebox{0.54}{\includegraphics[scale=1]{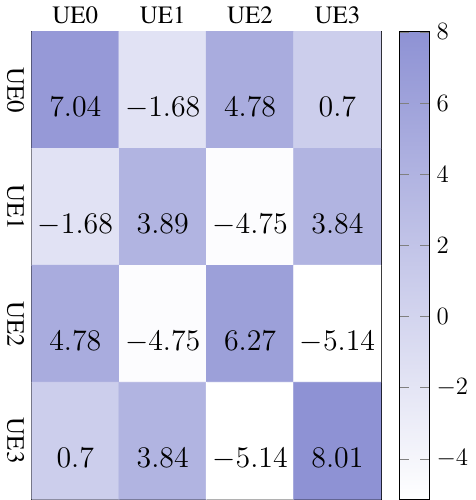}} \label{subfig:gramian-colocated}}
    \captionsetup[subfigure]{oneside,margin={0.85cm,0cm}}
	\hspace{-4mm}\subfloat[]{\hspace{0.5mm}\scalebox{0.45}{\includegraphics[scale=1]{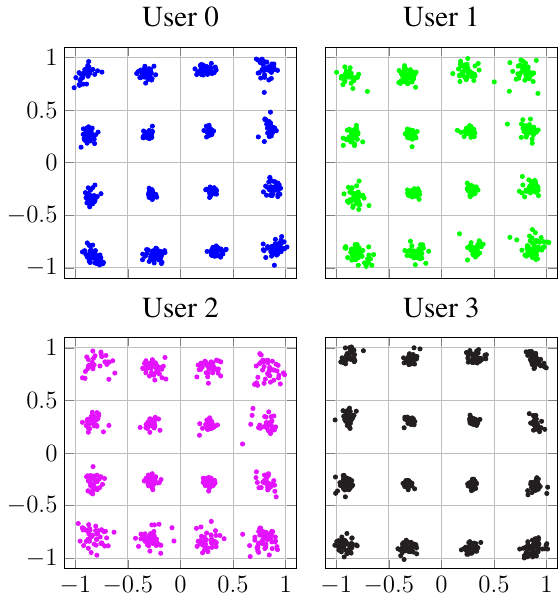}}\label{subfig:constellation-distributed}}
	\captionsetup[subfigure]{oneside,margin={0.85cm,0cm}}
	\hspace{-8.5mm}\subfloat[]{\hspace{8mm}\scalebox{0.54}{\includegraphics[scale=1]{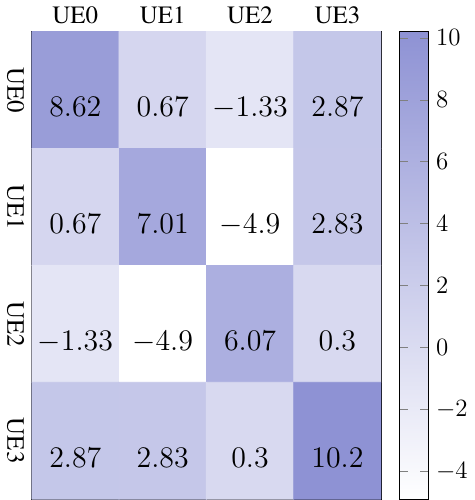}}\label{subfig:gramian-distributed}} 
	\caption{Results from one \gls{ofdm} symbol. User constellation after \gls{zf}-decoding and average Gram matrix magnitude (in dB) for (a), (b) --- co-located case and (c),(d) --- distributed case. The Gram matrix is averaged over the subcarriers. }	
    \label{fig:results}
\end{figure}
As proof-of-concept, we perform a real-time \gls{ul} transmission from the 4 single-antenna \glspl{ue} to the 256-antenna \gls{bs}, for the scenarios presented in Fig. \ref{fig:testbed}. A scenario where users are placed tightly in a square was chosen to validate the inter-user separation capabilities of the testbed. The users are sending frames continuously with visualisation happening in real-time. Fig. \ref{fig:results} show the equalized signal constellations and the observed Gram matrix, averaged over all the frequencies, of a received \gls{ofdm} data symbol for the 4 users, under the two deployment scenarios presented.

The results show a reduced inter-user interference in the distributed scenario, compared to the co-located one. 
Considering the \glspl{ue} placement,
the following conclusion can be drawn: for the co-located case, the beams resulted from \gls{mrc}, are sharper and more focused in depth, rather than the horizontal axis. Thus, the beams corresponding to the user sitting at the back of a square, will overlap with the ones sitting in the front, resulting in higher inter-user interference. This is clearly illustrated in Fig. \ref{subfig:gramian-colocated}, where the interference is worse for the user pairs (0,2) and (1,3), as compared to the Fig. \ref{subfig:gramian-distributed}, where the per-pair interference is diminished, as the beams are not only focused from one direction.
This was expected as the array aperture is increased, yielding better spatial resolution. 

\begin{figure}[t!]
	\centering
    \captionsetup[subfigure]{oneside,margin={0.85cm,0cm}}
	\hspace{-9.5mm}\subfloat[]{\hspace{8mm}\scalebox{0.485} 
    {\includegraphics[scale=1]{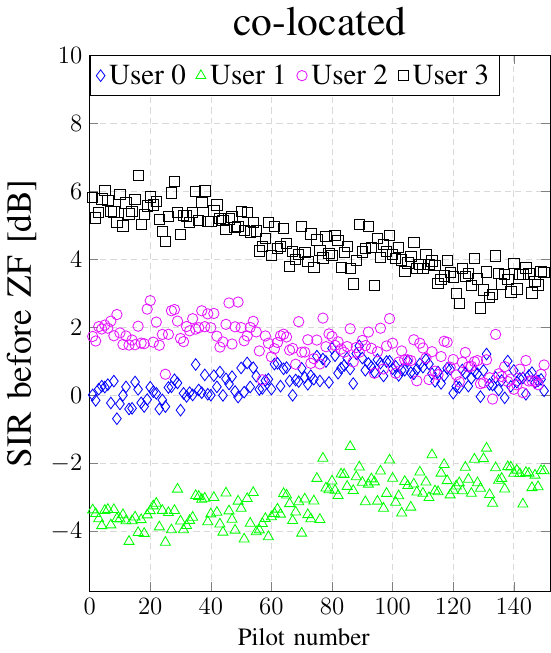}}\label{subfig:SIR-colocated}} 
	\captionsetup[subfigure]{oneside,margin={0.85cm,0cm}}
	\hspace{-3mm}\subfloat[]{\hspace{2mm}\scalebox{0.485}{\includegraphics[scale=1]{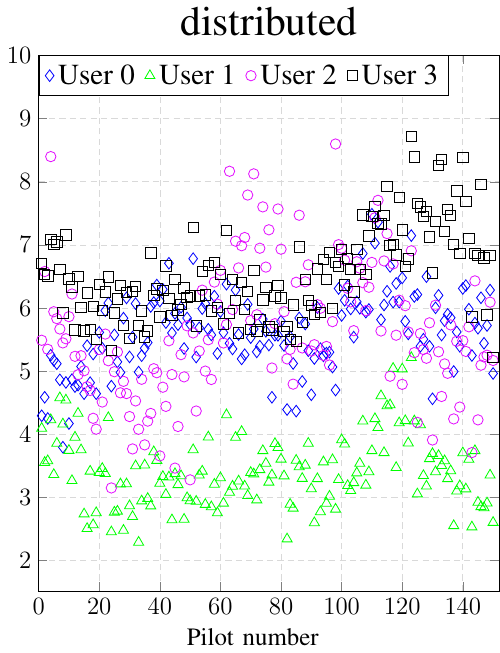}}\label{subfig:SIR-colocated}}
	\caption{SIR across subcarriers before \gls{zf}.}	
    \label{fig:sir}
\end{figure}

Fig. \ref{fig:sir}, where the observed \glspl{sir} before zero-forcing and across subcarriers are presented, strengthens the conclusion based on the Gram matrices, as it can be observed that there is more interference in the co-located case compared to the distributed. Nevertheless, the \gls{zf} equalization performs well in both cases, managing to cancel much of the interference (Fig. \ref{subfig:constellation-colocated} and Fig. \ref{subfig:constellation-distributed}). 

Table \ref{table:sir} further presents the average \gls{sir} for the cases where 64 and 128 antennas are used, showcasing the benefits with massive MIMO. Doubling the number of antennas(in the co-located case) leads to around 3 dB gain in the \gls{sir} for most cases. The tested configuration for fewer panels is marked in Fig. \ref{fig:testbed} with dotted blue and green squares, respectively.




\vspace{4mm}
\begin{table}[htbp]
\caption{\Gls{sir} before \gls{zf} and averaged over the frequencies.}
\resizebox{\columnwidth}{!}{%
\begin{tabular}{ccccc}
\hline
\multirow{2}{*}{\begin{tabular}[c]{@{}c@{}}Panel configuration \end{tabular}} & \multicolumn{4}{c}{Average SIR{[}dB{]}} \\
   & user 0 & user 1 & user 2 & user 3 \\ \hline
4 (co-located)  & -4.158 & -9.455 & -5.318 & -5.681 \\
8 (co-located) & -1.640 & -6.742 & -3.848 & -1.220 \\
16 (co-located) & 0.455  & -3.039 & 1.411  & 4.454  \\ 
16 (distributed) & 5.566  & 3.362 & 5.627  & 6.521  \\ \hline
\end{tabular}%
}
\label{table:sir}
\end{table} 


\section{Conclusion and future work}
\label{sec:Conclusion}

This worked presented a detailed solution and a proof-of-concept of a 256-antenna distributed coherent massive MIMO testbed implementation with real-time, fully-digital beamforming processing. To tackle the problems encountered with centralized processing, we design the processing nodes in a modular way, with a baseband distributed processing scheme that allows for lower data transfer between processing units. The testbed is easily scalable, as we employ a daisy-chain processing topology and the nodes do not require any hardware redesign when increasing the number of antennas.  
We perform an \gls{ul} transmission in both the co-located and distributed scenarios  and present the equalized user constellations as well as the Gram matrix and \gls{sir}, showing the performance of user separation and strengthening the theoretical assumptions about the benefits of distributed systems with a massive number of antennas.

Future work will focus on an optimized \gls{fpga} implementation of the zero-forcing algorithm and downlink functionality with reciprocity calibration, adding support for a higher number of users, and moving the synchronization of the distributed nodes to a white-rabbit based implementation. Moreover, the testbed can be used for acquiring measurements and further analyzing different distributed scenarios.



\section{Acknowledgement}

The authors would like to extend their gratitude to the AMD University Program for supplying the core hardware (ZCU216 RFSoCs) and whose general support has been instrumental in making design choices.
The authors would also like to thank Fredrik Tufvesson, Oscar Gustafsson for their suggestions, and Sirvan Abdollah Poor, Oscar Sanner, Erik Jonsson, Ivar Nilsson and Valentin Unger who supported the implementation of the testbed.

\balance
\bibliographystyle{IEEEtran}
\bibliography{references}

\end{document}